\documentclass{emulateapj}

\usepackage{amsmath}
\usepackage[]{natbib}

\newcommand{\etal}{{et al.~}}

\newcommand\hst{{\it HST}}

\slugcomment{Submitted for publication in {\sl The Astrophysical Journal}}

\shorttitle{Early-type galaxies in PEARS}
\shortauthors{Ferreras et al.}

\begin{document}

\title{Early-type galaxies in the PEARS survey:\\
 Probing the  stellar populations at moderate redshift}


\author{Ignacio Ferreras\altaffilmark{1,2},Anna Pasquali\altaffilmark{3},
Sangeeta Malhotra\altaffilmark{4}, James Rhoads\altaffilmark{4},
Seth Cohen\altaffilmark{4}, Rogier Windhorst\altaffilmark{4},
Nor Pirzkal\altaffilmark{5}, Norman Grogin\altaffilmark{5}, 
Anton M. Koekemoer\altaffilmark{5}, Thorsten Lisker\altaffilmark{6},
Nino Panagia\altaffilmark{5}, Emanuele Daddi\altaffilmark{7}, 
Nimish P. Hathi\altaffilmark{8}}
\altaffiltext{1}{ferreras@star.ucl.ac.uk}
\altaffiltext{2}{Mullard Space Science Laboratory, University College London, 
Holmbury St Mary, Dorking, Surrey RH5 6NT, UK.}
\altaffiltext{3}{Max-Planck-Institut f\"ur Astronomie, Koenigstuhl 17, D-69117
Heidelberg, Germany}
\altaffiltext{4}{Department of Physics and Astronomy, 
Arizona State University, P.O. Box 871504, Tempe, 
AZ 85287-1504}
\altaffiltext{5}{Space Telescope Science Institute, 3700 San Martin Drive, 
Baltimore, MD 21218}
\altaffiltext{6}{Astronomisches Rechen-Institut, Zentrum f\"ur Astronomie, Universit\"at 
Heidelberg, M\"onchhofstr. 12-14, D-69120 Heidelberg, Germany}
\altaffiltext{7}{CEA Saclay, Orme des Merisiers, 91191 Gif-sur-Yvette 
Cedex, France}
\altaffiltext{8}{Department of Physics and Astronomy, University of 
California, Riverside, CA 92521}

\begin{abstract}
Using \hst/ACS slitless grism spectra from the PEARS program, we study
the stellar populations of morphologically selected early-type
galaxies in the GOODS North and South fields. The sample -- extracted
from a visual classification of the (v2.0) \hst/ACS images and
restricted to redshifts z$>$0.4 -- comprises 228 galaxies ($i_{\rm
F775W}<24$~mag, AB) out to z$\lesssim$1.3 over 320 arcmin$^2$, with a
median redshift z$_{\rm M}$=0.75. This work significantly increases
our previous sample from the GRAPES survey in the HUDF ($18$ galaxies
over $\sim$11~arcmin$^2$; Pasquali et al. 2006b).  The grism data allow us
to separate the sample into `red' and `blue' spectra, with the latter
comprising 15\% of the total. Three different grids of models
parameterising the star formation history are used to fit the
low-resolution spectra. Over the redshift range of the sample --
corresponding to a cosmic age between 5 and 10 Gyr -- we find a strong
correlation between stellar mass and average age, whereas the {\sl
spread} of ages (defined by the RMS of the distribution) is roughly
$\sim 1$~Gyr and independent of stellar mass. The best-fit parameters
suggest it is formation epoch and not formation timescale, that best
correlates with mass in early-type galaxies.  This result -- along
with the recently observed lack of evolution of the number density of
massive galaxies -- motivates the need for a channel of (massive)
galaxy formation bypassing any phase in the blue cloud, as suggested
by the simulations of Dekel et al. (2009).
\end{abstract}


%%%%%%%%%%%%%%%%%%%%%%%%%%%%%%%%%%%%%%%
\section{Introduction}

For nearly a century, we have known that the local galaxy population follows
a bimodal colour distribution, with a majority of star-forming
galaxies and a minority of early-type systems
\citep{edwin,strat01,blan03}.  This distribution \citep[which has been
observed up to z$\sim$1,][]{Bell04} is commonly interpreted as the
outcome of different star-formation histories, with blue galaxies
undergoing a much more prolonged star-formation activity than the
early-types. This does not rule out that early-type galaxies may also
experience occasional episodes of star formation at z$\sim$0.
However, if they do, they produce an amount (in mass) of young stars
definitively lower than the blue-peak galaxies
\citep{sct00,pca,ben07}.  Early-type galaxies are known to exhibit a
tight colour-magnitude relation \citep[up to z$\lesssim$1,][]{sed98}
which is the direct result of a metallicity sequence where more
massive galaxies are also metal-richer \citep{ble92,ka97,vaz01,ber03}.
The small scatter in their colours and mass-to-light ratios \citep[see
e.g. ][]{kel00}, and their relatively high [$\alpha$/Fe]
\citep{nel05,thom05} suggest that early-type galaxies formed the bulk
of their stars at high redshift \citep[z$\gtrsim$3][]{vdk98,thom05}
and on a relatively short timescale \citep[i.e. $\sim$1 Gyr for the
more massive galaxies,][]{thom99}. While these properties can be well
reproduced by the ``monolithic collapse hypothesis''
\citep{els62,lar74}, this scenario does not fit well within the
currently accepted $\Lambda$CDM galaxy formation paradigm, in which 
galaxies form through hierarchical buildup. In semi-analytical models,
the hierarchical growth of dark matter structures is complemented by
the addition of simple prescriptions of the baryonic physics which
describe star formation, chemical enrichment, stellar and AGN feedback
\citep[e.g.][]{evan05,crot06,hop06,mon07,hop08,som08}. The model
predictions are in reasonable agreement with the observed properties
of early-type galaxies \citep[see e.g. ][]{kav05,Delu06,ks06a,ks06b},
but they still can not reproduce the observed high [$\alpha$/Fe]
ratios.  On the other hand, support to the merging scenario is given
by the detection, at rest-frame UV wavelengths, of recent episodes of
star formation in early-type galaxies
\citep{fs00,kav07,ben07,scha07,kav08} and by the evidence that less
massive early-type galaxies increased their mass by 20 -- 40\% since
z=1 \citep{Bell04,fl05,Treu05a}

We note that all the knowledge collected on early-types is primarily
based on the nearby galaxy population. Extending the study to a large
number of early-type galaxies at intermediate and high redshifts
allows us to trace their mass-assembly and star formation histories in
great detail and, consequently, to fine-tune the theory of galaxy
formation and evolution.  Ferreras et al. (2009a,b) derived the
evolution of number density and size (i.e. half light radius) for a
sample of early-type galaxies between z=1 and z=0, extracted from the
GOODS North and South fields \citep[see also][]{fl05} and compared it
with recent surveys such as GOODS/MUSIC \citep{fon06}, DEEP2
\citep{con07,truj07}, COMBO17 \cite{borch06}, 2dFGRS \citep{crot05}
and SDSS \citep{shen03}.  They found that massive early-types do not
exhibit any significant change in comoving number density over the
past 8 billion years.  On the contrary, there exists a noticeable
evolution in size, implying an increase of about a factor of 2 for
early-type galaxies more massive than $10^{11}$M$_{\odot}$ and a
factor of $\sim$50\% for the less massive ones. This result is in
agreement with \citet{daddi05}, \citet{con07}, Trujillo \etal
(2006,2007), \citet{sca07}, \citet{vdk08} and \citet{vdwel08}. The
latter also found that the $\sigma$--R$_e$ relation is different for
nearby and distant early-type galaxies, suggesting a significant
change in the dynamics of these galaxies. 
Given the exposure times needed for accurate velocity dispersions
in these galaxies, it does not come as a surprise that the amount of
evolution of the velocity dispersion is still controversial \citep[see
e.g.][]{ct09,vdk09,cap09}.
Comparison with the semi-analytical models of
Khochfar \& Silk (2006a,b) -- where galaxy growth occurs via
dissipative major mergers -- suggests that the constant comoving
number density at high $M_*$ is due to a balance between the ``sink''
(i.e. loss due to mergers of massive galaxies producing more massive
galaxies) and the ``source'' terms (i.e. gain from mergers at lower
mass) in the 0$<$z $<$1 range. \citet{fl09a} also showed that the
slope of the Kormendy relation in the 0.4$<$z$<$1.4 range is
consistent with z=0 values \citep[see also][]{zieg99,wad02}. Only the
average surface brightness is seen to change, according to pure
passive evolution of old stellar populations.

The fact that stellar populations in early-type galaxies undergo
passive evolution is confirmed by their colours and photometric
spectral energy distributions (SEDs). Morphologically-selected
early-type galaxies have been so far identified in clusters up to
z$\simeq$1.4 and their colours have been proven to be consistent with
those of z=0 early-types through pure passive evolution of their stars
\citep[see e.g.][]{dres90,as93,rs95,vdm07,cool08,whil08}. 
\citet{rse97}, \citet{sed98} and \citet{blk03} showed that early-type
galaxies in clusters at z$\sim$0.5 -- 1.2 complete their star
formation by z$\geq$3 \citep[see also
][]{fcs99,glad98,kod98,nel01,vdk00}. The same can be said for the most
massive early-type galaxies ($M_* > 10^{11} M_{\odot}$) in the field
at the same redshift, while the less massive ones appear to sustain a
more prolonged star formation \citep[see
e.g.][]{Bell04,vdwel04,vdwel05,vdk03,Treu05a,Treu05b}.

Although photometric colours and SEDs are clearly the most affordable
observables to obtain in terms of telescope time, they do not
disentangle between age and metallicity effects as accurately as
spectra and line indices. The measurement of the Mg$b$ index and of
the $H_{\delta}$ and H$_{\gamma}$ strength in early-type galaxies up
to z=0.83 has independently provided z$>$2 -- 3 as the redshift at
which their stars formed \citep{zieg97,kel01}. Spectroscopy from the
ground has made also possible to use a set of FeII, MgII and MgI lines
in the rest-frame UV and identify massive early-type galaxies ($M_* >
10^{11} M_{\odot}$) up to z$\simeq$2.15
\citep{cim04,glaz04,McCa04,sar05}. In particular, \citet{cim04}
stacked together the rest-frame UV spectra of 4 early-type galaxies
with $1.6 < z < 1.9$ and compared the average with synthetic spectra
of simple stellar populations to find a mean stellar age of about 1
Gyr for solar metallicity (or better 0.5 $<$ age $<$ 1.5 Gyr for 2.5
$> Z/Z_{\odot} >$ 0.4). Similarly, \citet{cim08} combined the spectra
of 13 early-type galaxies at $\langle$z$\rangle$=1.6 and determined a
mean stellar age in the range 0.7 - 2.8 Gyr for $Z =
1.5-0.2Z_{\odot}$, corresponding to a formation redshift z$_{\rm
F}>$2.  Along this line, Kriek et al. (2008) averaged the rest-frame
optical spectra of 16 early-type galaxies at 2$<$z$<$3 to find the
typical SED of a poststarburst galaxy with an age of about 1 Gyr for
solar metallicity. They concluded that early-type galaxies are likely
to form between z$_{\rm F}$=2 and 3 and are not yet in place at z=3
\citep[see also][]{dun96,hy97,driv98,bram07,kod07}.

At redshifts z$\gtrsim$1, though, the performances of ground-based
spectrographs is severely limited by the atmosphere and its OH lines,
and also by the optical faintness of the targets ($I\gtrsim$ 23~mag).
The high-redshift population of early-type galaxies is, thus, best
investigated from space with low-resolution spectroscopy. This is the
case of the \hst\ Advanced Camera for Surveys (ACS), which is equipped
with a slitless grism covering the spectral range 5500 - 10500 \AA\/
with a dispersion of 40 \AA/pix in the first spectral order
\citep{ap06a}.  The \hst/ACS grism has been used for spectroscopic
surveys like GRAPES (Pirzkal et al. 2004) and PEARS (Malhotra et al.,
in preparation), where early-type galaxies brighter than $I \simeq 26$~mag
have been identified up to z=2.5 and their individual spectra
(typically with S/N $>$ 10 pixel$^{-1}$) have been analyzed to
constrain their stellar populations. The low resolution of these
spectra clearly prevent us from measuring line indices, but allows us
to detect the broad feature Mg$_{UV}$ in the rest-frame 2640--2850
\AA\/ \citep[][]{daddi05,mar06} at z$>$1.3 and measure the 4000\AA\/
break at 0.4$<$z$<$1.2 and use it along with the continuum to estimate
stellar ages and metallicities down to M$_*\simeq
10^9-10^{10}$M$_{\odot}$ \citep{ap06b}. This limiting stellar mass is
a factor of about 10 lower than what can be achieved with ground-based
spectroscopy, and makes it possible, for the first time, to study the
stellar populations at the faint end of the red sequence at z$\sim$1.
For example, \citet{ap06b} identified 18 early-types in the Hubble
Ultra Deep Field \citep{sb06} with $0.49 \leq z \leq 1.15$. They
fitted the individual \hst/ACS grism spectra with synthetic star
formation histories, based on the models of \citet{bc03} and including
chemical enrichment \citep{fs03}.  They found stellar ages varying
from 3--4 Gyr at z$>$0.65 to 5--6~Gyr at lower redshifts, stellar
metallicities between 1 and 0.03 $Z_{\odot}$ and stellar masses
between $3 \times 10^9$ and $3 \times 10^{11} M_{\odot}$. The
estimated ages are consistent with a formation redshift z$>$2--5.

\begin{figure}
\epsscale{1.2}
\plotone{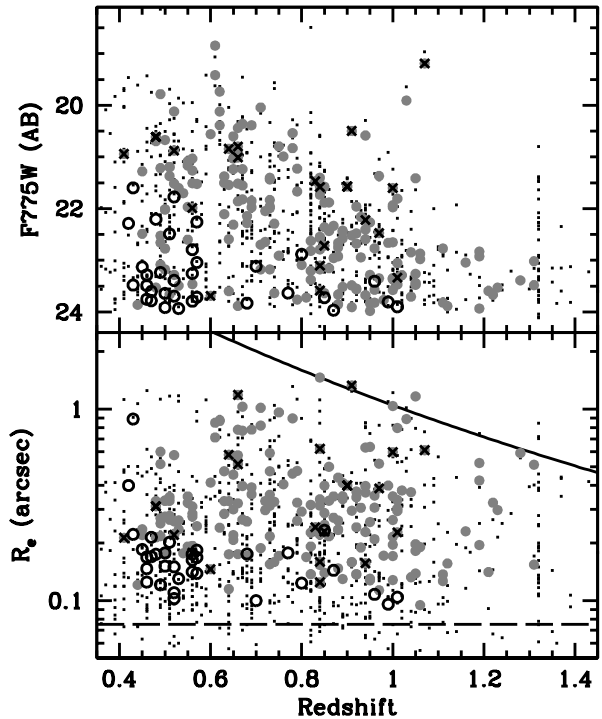}
\caption{Observed properties of the PEARS early-type galaxy sample,
extracted from visual classification in GOODS N and S \hst/ACS images
down to $i_{\rm AB}=24$~mag \citep[full sample shown as dots,][]{fl09a}.
We cross-correlate the total sample of 377 (CDFS) and 533 (HDFN)
early-types with the PEARS extractions. After visual rejection of bad
spectra and rejection of redshfts z$<$0.4 we end up with the sample
presented in this figure (circles): 91+13 (CDFS) and 104+20 (HDFN)
galaxies. The first number represents those with a ``red'' SED (grey
filled circles): prominent 4000-break and no emission lines. The
second number corresponds to ``blue'' SEDs (open circles, see text for
details). We overlay an 'X' symbol on those galaxies with an X-ray
detection from the Chandra deep fields \citep{alex03,luo08}. Notice
that all X-ray detections correspond to ``red'' galaxies. The dashed line
in the bottom panel illustrates the resolution limit at 
2R$_e\sim$~FWHM$_{\rm HST/ACS}$.
The solid line is the surface brightness limit, considering the cosmological
dimming, at $i_{\rm AB}=24$~mag (AB), assuming a surface brightness at z=0.7
of $\mu_{\rm F775W}=22$~mag/arcsec$^2$.
\label{fig:Obs}}
\end{figure}

In this paper, we extend the analysis of \citet{ap06b} to a sample of
early-type galaxies that is a factor of 11 larger than in the
HUDF. The objects -- 228 in total -- are extracted from the sample of
morphologically selected early-type galaxies of \citet{fl09a} in the
GOODS North and South fields and have been observed as part of
PEARS\footnote[1]{\tt http://archive.stsci.edu/prepds/pears/}. 
This is the largest database of individual, high S/N grism
spectra obtained so far for early-type galaxies at z$\sim$1, 
that allows a study of their stellar populations as a function of
stellar mass and redshift, and can place more stringent constrains on
the formation and evolution of (not only massive) early type galaxies
since z$\sim$1.

Throughout this paper a standard $\Lambda$CDM cosmology from the
WMAP-year 5 data is used to determine ages and rest-frame properties
\citep{wmap5}.

%%%%%%%%%%%%%%%%%%%%%%%%%%%%%%%%%%%%%%%%%%%%%
%%%%%%%%%%%%%%%%%%%%%%%%%%%%%%%%%%%%%%%%%%%%%
%%%%%%%%%%%%%    TABLE 1    %%%%%%%%%%%%%%%%%
%%%%%%%%%%%%%%%%%%%%%%%%%%%%%%%%%%%%%%%%%%%%%
%%%%%%%%%%%%%%%%%%%%%%%%%%%%%%%%%%%%%%%%%%%%%

\begin{deluxetable*}{rrrrrrc}
\tabletypesize{\scriptsize}
%\rotate
\tablecaption{Catalog of PEARS early-type galaxies: Observed data
\label{tab:catobs}}
\tablewidth{0pt}
\tablehead{
\colhead{PID$^a$} & \colhead{RA} & \colhead{Dec} & \colhead{F775W} & \colhead{R$_e$}
& \colhead{z} & \colhead{R/B$^b$}\\
\colhead{} & \multicolumn{2}{c}{J2000, deg.} & \colhead{(AB)} & \colhead{pix$^c$} & \colhead{} &
\colhead{}}
\startdata
  65620 &  53.1649933 & $-$27.819334 & 21.37 & 17.33 & 0.97 & R\\
  56798 & 189.1671143 & $+$62.218311 & 20.70 & 11.04 & 0.48 & R\\
  83499 & 189.1973572 & $+$62.274567 & 21.57 & 13.30 & 0.85 & R\\
  75762 & 189.1127472 & $+$62.252712 & 21.23 & 15.86 & 0.79 & R\\
  61447 &  53.1852264 & $-$27.827837 & 23.10 &  4.16 & 0.84 & R\\
 127697 &  53.0613937 & $-$27.698137 & 22.53 &  7.25 & 0.49 & R\\
  52017 & 189.1557922 & $+$62.208328 & 23.35 &  3.78 & 0.97 & R\\ 
  61235 &  53.1881218 & $-$27.827766 & 22.87 &  6.35 & 1.06 & R\\
  69419 &  53.1323776 & $-$27.814238 & 22.91 &  5.72 & 0.75 & R\\
%\hline
  40498 & 189.1528931 & $+$62.186619 & 23.72 &  7.77 & 0.85 & B\\
  47252 & 189.2056274 & $+$62.198719 & 23.41 &  3.60 & 0.96 & B\\
  42729 &  53.2094307 & $-$27.861725 & 23.79 &  5.44 & 0.56 & B\\
  17362 &  53.1624603 & $-$27.914028 & 22.49 &  6.72 & 0.51 & B\\
  57554 & 189.2057190 & $+$62.219887 & 22.21 &  5.83 & 0.48 & B\\
 119723 & 189.3504181 & $+$62.327610 & 22.89 &  4.12 & 0.80 & B\\
  53462 &  53.1896324 & $-$27.841940 & 22.79 &  5.84 & 0.56 & B\\
 107477 &  53.0674744 & $-$27.739756 & 22.29 & 13.32 & 0.42 & B\\
 104645 & 189.3910980 & $+$62.301193 & 23.78 &  5.66 & 0.47 & B\\
\enddata
\tablecomments{A full version of table \ref{tab:catobs} is published in its 
entirety in the electronic edition of the {\it Astrophysical Journal}. 
A portion is shown here for guidance regarding its form and content.}
\tablenotetext{a}{PEARS ID number.}
\tablenotetext{b}{`Red' vs `Blue' early-type (see text for details).}
\tablenotetext{c}{1 pixel = 0.03 arcsec}
\end{deluxetable*}
%%%%%%%%%%%%%%%%%%%%%%%%%%%%%%%%%%%%%%%

\begin{figure}
\epsscale{1.2}
\plotone{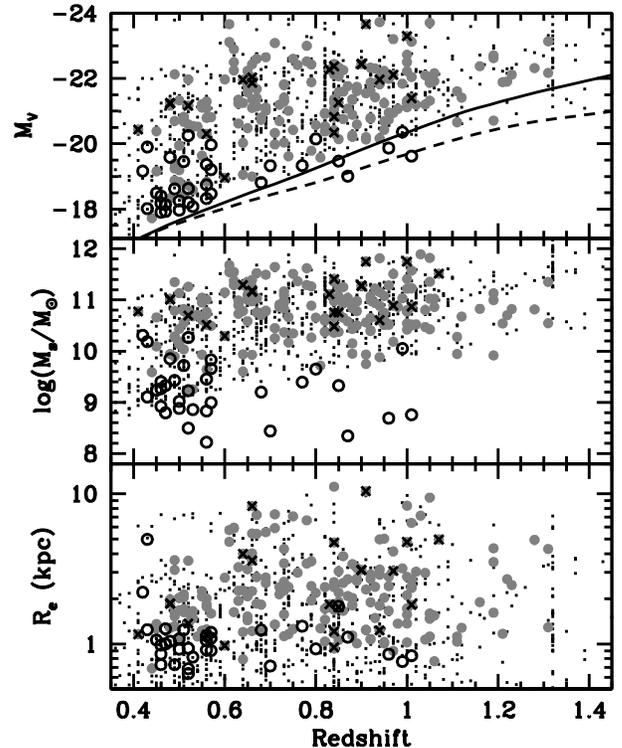}
\caption{Rest-frame properties of PEARS early-type galaxies. The
sample is shown with respect to absolute V-band magnitude (top),
stellar mass (middle) and projected half-light radius (bottom). The
masses and magnitudes are computed using the best-fit SFH. 'Red' and
'blue' galaxies are represented by filled and open circles,
respectively. The solid lines at the top panel track the $i_{\rm
AB}=24$~mag cut in apparent magnitude for two extreme cases: an old,
passively evolving population (solid) or an extended star formation
history (dashed). The small dots correspond to the original sample of
visually-classified spheroidal galaxies from \citet{fl09a}.  We
overlay an 'X' symbol on those galaxies with an X-ray detection from
the {\sl Chandra} deep fields North and South \citep{alex03,luo08}.
\label{fig:Phys1}}
\end{figure}

%%%%%%%%%%%%%%%%%%%%%%%%%%%%%%%%%%%%%%%
\section{Collecting the sample of early-type galaxies}

The data consist of ACS images taken with the WFC G800L grism. The
PEARS survey comprises eight pointings towards the GOODS North and
South fields -- 4 pointings each -- with three \hst\ roll angles
for each pointing (20 \hst\ orbits per field), plus the HUDF with
four roll angles (40 \hst\ orbits). A forthcoming paper (Malhotra
et al. 2009, in preparation) will describe the PEARS project and data
in detail. The HUDF data comes from an earlier similar project, GRAPES
\citep{nor04,ap06b}. We use the PEARS IDs from the master catalogue of
source extractions. The G800L grism produces slitless, low-resolution
spectra in the wavelength range 6000--9500\AA. The optimal spectral
resolution ($R\sim 100 $) is achieved for unresolved sources, whereas
for our sample of early-type galaxies, the effective resolution is
lower, around $R\sim 50$.

The master catalogue of PEARS detections was cross-correlated with the
sample of 910 visually-classified early-type galaxies in the GOODS
North and South fields of \cite{fl09a}. This sample completes the one
presented previously in the CDFS field \citep{fl05}. The selection
consists of visual inspection (in all four ACS bands) of all source
detections brighter than $i_{\rm F775W}<24$~mag (AB) by four people,
with two rounds of visual inspection for those galaxies that were
harder to classify. No colour cuts or CAS-based selections were made,
although the sample is consistent with the high concentration and low
asymmetry expected in spheroidal galaxies with respect to later
morphological types (Lisker et al., in preparation). Furthermore, the
visually-classified sample comprises 80\% of red, quiescent stellar
populations. We believe visual inspection alone is much more efficient
at selecting spheroidal galaxies, eliminating contamination from
late-type galaxies or unresolved sources, and removing the bias
inherent in color-based selection methods.  From this catalogue we
extract the available PEARS spectra of each identified object from
our database. The spectra are background corrected and scaled to match the
direct image magnitudes of each object.  Each file contains the
spectra extracted for the different PAs (typically around three 
PAs per galaxy). We compute a simple average of the spectra available for
each object, possibly excluding those PAs whose spectra are clearly
deviant.  The errors in the final combined spectra are computed by
propagating the flux errors in each individual SED. The spectra were
visually inspected, and some of them had to be discarded because of a
very low signal-to-noise ratio, contamination from nearby sources, or
because they were located close to the edge of the field. Out of the
original sample of 910 galaxies (533 in HDFN and 377 in CDFS) we
extracted 136 spectra in the North and 131 in the South. The number of
rejected galaxies is 19 and 18 in the North and South fields,
respectively.

Regarding redshift, we start with estimates from the PEARS-based
general catalog of photometric redshifts (Cohen et al. in
preparation), and the photometric redshift catalogue of \cite{mob04}
and \cite{fl09a}. Spectroscopic redshifts are available from FORS2
\citep{vanz08} and the VIRMOS-VLT Deep Survey \citep[VVDS,][]{pop08}
in the South, and from the Team Keck Redshift Survey
\citep[TKRS,][]{tkrs04} in the North. However, our slitless grism data
has enough S/N to allow for an estimate of the redshifts {\sl
directly} from the low-resolution spectra. Two of us (I.F. and A.P.)
inspected visually each SED and compared it with a number of template
spectra to determine a rough estimate of the redshift. Then we ran a
simple code that uses the guess and templates to accurately determine
the redshift via a standard maximum likelihood method. The accuracy of
our redshifts is as good as spectroscopic estimates for most of the
galaxies, especially those with a prominent 4000\AA\ break (i.e. the
majority of the sample) or those with emission lines. Incidentally,
some of our (grism-based) redshift estimates {\sl improved} on the
spectroscopic values from pipeline-processed databases such as
GOODS/FORS2. This improvement is to be expected for the faintest
galaxies with a quiescent spectra. In those cases the continuum from
the ground-based spectroscopic data is very noisy and, simply put, our
low-resolution spectra yield more photons per pixel over the CCD,
significantly improving the signal. Hence, we consider our quoted
redshifts as spectroscopically accurate.

Finally, we apply a cut in redshift, removing from the analysis all
galaxies with redshift z$<$0.4. At those redshifts the 4000\AA\ break
-- on which the modelling strongly relies -- falls outside of the
sensitivity region of the G800L grism. This cut reduces the sample to
124 and 104 galaxies in the North and South fields, bringing the total
to 228 early-type galaxies with z$>$0.4 and $i_{\rm F775W}<24$~mag
(AB).  The median redshift of the sample is z$_{\rm M}$=0.75. If we
measure the signal-to-noise ratio as the average S/N per pixel within
the spectral window 8000-8500\AA\ , our sample has a median value of
S/N$\sim 15$ per pixel.

\begin{figure}
\epsscale{1.2}
\plotone{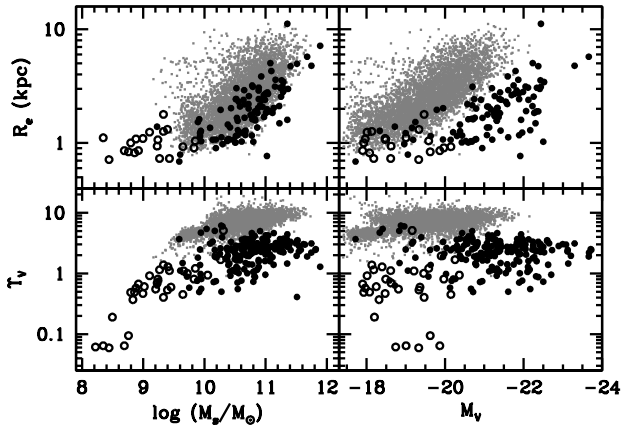}
\caption{Comparison of half-light radius (R$_e$; {\sl top}) and
$V$-band M/L ($\Upsilon_V$ with respect to the solar value; 
{\sl bottom}) with a z$\sim 0$ sample of galaxies with Sersic index
$n_S>2.5$, extracted from the SDSS NYU Value Added Galaxy Catalogue
\citep[grey dots,][]{blan05}. The half-light radius ({\sl top}) and
M/L in the $V$ band is shown as a function of stellar mass ({\sl
left}) or absolute magnitude.`Red' (`blue') PEARS galaxies are
represented as black solid (open) circles.\label{fig:Phys2}}
\end{figure}

\subsection{Sample Properties}

The main observational data are presented in
table~\ref{tab:catobs}. Figure~\ref{fig:Obs} shows the distribution of
total apparent magnitude ({\sl top}) and size (half-light radius; {\sl
bottom}).  The original sample of visually-classified galaxies from
\citet{fl09a} is shown as small dots. The PEARS sample of early-type
galaxies is shown as circles. We subdivide the sample into `red' (grey
dots) and `blue' types (open circles). This classification is taken
from the original data and is determined by the spectral template that
gives the best fit in the analysis of photometric redshifts. For the
PEARS sample presented here, the division is much more accurate because
this information is complemented by the visual appearance of the
low-resolution SED. `Red' galaxies feature prominent 4000\AA\ breaks,
whereas `blue' galaxies have a clear blue continuum with the
occassional presence of emission lines. The sample comprises 195 `red'
and 33 `blue' spheroidal galaxies. The accuracy of the total
magnitudes and half-light radii is estimated from simulations of
synthetic galaxies with Sersic profiles with similar values of size
and magnitude as the original sample. We recover the original values
with an uncertainty $\Delta i_{\rm F775W}=0.05$~mag and
$\Delta$R$_e/$R$_e=0.09$ \citep{fl09a}.  We overlay 'X' symbols over
those galaxies with an X-ray detection from the {\sl Chandra} Deep fields
North and South \citep{alex03,luo08}.  22 sources have an X-ray
detection which extends over the full range in apparent magnitude. It
is worth emphasizing that all X-ray detected sources in our sample are
photometrically classified as `red' galaxies.

\begin{figure}
\epsscale{1.2}
\plotone{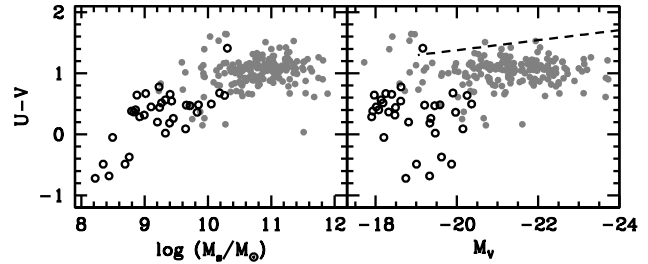}
\caption{Rest-frame $U-V$ color-magnitude relation ({\sl right})
and color-stellar mass relation ({\sl left}). Filled (open) circles
correspond to photometric `red' (`blue') galaxies. The dashed line
gives the z$\sim 0$ colour-magnitude relation of Coma early-type
galaxies from \citet{ble92}.
\label{fig:CMR}}
\end{figure}

The distribution of rest-frame properties of our sample is shown in
figure~\ref{fig:Phys1}, where the absolute magnitude in the $V$ band
(top); stellar mass (middle) and physical half-light radius (bottom)
is shown with respect to redshift. The K-correction and the M/L
estimate needed for the top two panels is taken from the best fit
models of the runs explained in \S3.  The `red' (`blue') galaxies are
shown as filled grey (open black) circles. `Blue' early-types are
mostly faint, low-mass galaxies. We should emphasize that our
selection process (visual classification) does not introduce any bias
regarding the color distribution. The lines in the top panel
correspond to the $i_{\rm F775W}<24$~mag limit imposed on the original
sample. Since the translation from apparent to absolute magnitude
depend on the template SED for the K-correction, we show the apparent
magnitude limit corresponding to two types of populations: quiescent
(solid line) and young (dashed line). These two cases are the extrema
of the photometric types used in \citet{fl09a}, namely t=0 and
t=7. These types correspond to an exponential star formation history
at fixed solar metallicity, beginning at a formation redshift z$_{\rm
F}=3$ with timescale $\tau=0.05$~Gyr (t=0) or $\tau=8$~Gyr (t=7).
Similarly to figure~\ref{fig:Obs}, the 'X' symbols are overlaid on
those galaxies with an X-ray detection from the {\sl Chandra} Deep
Fields North and South \citep{alex03,luo08}.  Almost all X-ray
detections correspond to galaxies brighter than $M_V<-20$, with
stellar masses above $10^{10}$M$_\odot$. However, many of them are
concentrated, with half-light radii $R_e\sim 1$~kpc.

Figure~\ref{fig:Phys2} shows the physical half-light radius (R$_e$;
{\sl top}) and M/L in the $V$ band ($\Upsilon_V$ with respect to the
solar value; {\sl bottom}) as a
function of stellar mass ({\sl left}) or absolute magnitude ({\sl
right}). Black filled (open) circles correspond to `red' (`blue')
galaxies.  In order to compare with z=0 galaxies, we include as grey
dots the values from the DR4 NYU Value Added galaxy catalogue
\citep{blan05} with a Sersic index $n_S>2.5$.  Local galaxies appear
significanty larger and with higher M/L. The latter can be explained
by pure passive evolution of the stellar populations, but the change
in size requires a mechanism beyond stellar evolution.  This result is
shown for the full sample of GOODS spheroidal galaxies in
\citet{fl09b} and confirms previous results at similar or higher
redshifts \citep{truj07}. Size evolution is even stronger when
galaxies at z$\sim$2 are considered \citep{vdk08,bui08}.
Recent attempts at explaining the size evolution invoke a 
significant ejection of gas mass caused by quasar feedback \citep{fan08},
or via dry mergers in the standard paradigm of hierarchical evolution
\citep{hop09}.

\begin{figure}
\epsscale{1.2}
\plotone{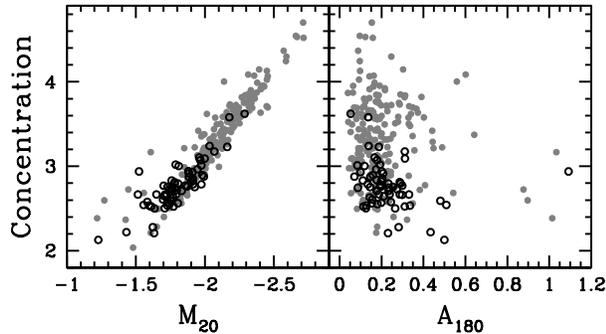}
\caption{Quantitative morphology of the PEARS early-type galaxy
sample. The plot shows the concentration, M$_{20}$ (a normalized
second order moment of the 20\% brightest pixels) and
asymmetry. Filled grey (black open) circles correspond to `red'
(`blue') spheroidal galaxies. See text for details.\label{fig:cas}}
\end{figure}

The rest-frame $U-V$ color-magnitude and color-stellar mass relations
are presented in figure~\ref{fig:CMR}, with the same coding for
photometric type as in the previous figures. Notice color is
correlated more strongly with stellar mass than with absolute
magnitude.  The dashed line gives the Coma color-magnitude
relation from \cite{ble92} for E+S0s. The extent of the line covers
the magnitude range probed in Coma. The relation at high redshift is
in good agreement with Coma if we consider that passive evolution of a
stellar population with solar metallicity, formed at z$_{\rm F}=3$
changes its U$-$V colour by $\sim 0.3$~mag between the median redshift
of our sample $\langle$z$\rangle\sim 0.7$ and z$=0$. Instead of
dealing with simple, K-corrected colors, we use the spectroscopic data
to determine the best fit ages from models of stellar populations.
In the next section we compare the low-resolution SEDs from PEARS with
a large grid of star formation histories to transform this information
into a more meaningful diagram involving stellar ages.

A rough morphological quantification of the sample is presented in
figure~\ref{fig:cas}, where the concentration, M$_{20}$ and asymmetry
values are shown for the $i_{\rm F775W}$ images (Lisker \etal in preparation). Concentration
\citep[see e.g.][]{Ber00} is defined as C$=5\log$R$_{80}$/R$_{20}$,
where R$_X$ is defined as the radius within which the enclosed flux
corresponds to X \% of the total flux (as measured within the Petrosian
radius). M$_{20}$ is defined as the normalized second order moment of
the brightest 20\% of the pixels no matter where they are
\citep{lotz04}. The asymmetry parameter (A$_{180}$) is obtained by
subtracting a 180$^{\rm o}$ rotated image from the original.
A$_{180}$ is given by this difference -- summed and expressed as a
fraction of the original flux \citep{con03}.  The figure shows that
most galaxies (whose selection was purely based on a visual
classification, i.e. NOT extracted from their CAS values) are highly
concentrated both in C and M$_{20}$ and have low asymmetry. Notice the
blue photometric types (open circles) coincide with the lowest values of
concentration, although those values are still high for a general
population involving all galaxy morphologies, which can have
concentrations as low as C$\sim$1.5. The asymmetry of blue and red
early-types is similar.

%%%%%%%%%%%%%%%%%%%%%%%%%%%%%%%%%%%%%%%%%%%%%
%%%%%%%%%%%%%%%%%%%%%%%%%%%%%%%%%%%%%%%%%%%%%
%%%%%%%%%%%%%    TABLE 2    %%%%%%%%%%%%%%%%%
%%%%%%%%%%%%%%%%%%%%%%%%%%%%%%%%%%%%%%%%%%%%%
%%%%%%%%%%%%%%%%%%%%%%%%%%%%%%%%%%%%%%%%%%%%%

\begin{deluxetable}{lccc}
\tabletypesize{\scriptsize}
%\rotate
\tablecaption{Range of parameters explored for the three models
considered in this paper (see text for details).
\label{tab:params}}
\tablewidth{9cm}
\tablehead{
\colhead{Model/Param} & \colhead{MIN} & \colhead{MAX} & Comments}
\startdata
SSP & & 2 params \\
$t$ & 0.1 Gyr & z$_{\rm F}=10^a$ & Age\\
$\log(Z/Z_\odot)$ & -1.5 & +0.3 & Metallicity\\
\hline
EXP  & & & 3 params \\
$\log\tau$ (Gyr) & -1 & +0.6 & Exp. Timescale\\
z$_{\rm F}$ & z$-0.1$~Gyr$^b$ & 10 & Formation epoch\\
$\log(Z/Z_\odot)$ & -1.5 & +0.3 & Metallicity\\
\hline
2BST & & & 4 params\\
$t_O$ & 2   & z$_{\rm F}=10^a$  & Old (Gyr)\\
$t_Y$ & 0.1 &  2  & Young (Gyr)\\
$f_Y$ & 0.0 & 1.0 & Mass fraction\\
$\log(Z/Z_\odot)$ & -1.5 & +0.3 & Metallicity\\
\hline
CSP  & & & 3 (free) params\\
$\nu$ (Gyr$^{-1}$) & \multicolumn{2}{c}{20} & SF Efficiency (fixed)\\
$\tau_f$ (Gyr) & -1 & +0.3 & Gas Infall timescale\\
$\beta$ (Gyr) & 0 & 1 & Outflow fraction\\
z$_{\rm F}$ & z$-0.1$~Gyr$^a$ & 10 & Formation epoch\\
\enddata
%\tablecomments{None}
\tablenotetext{a}{SSP,2BST: The oldest age possible for the old component
corresponds to a formation redshift z$_{\rm F}=10$ at the redshift of the galaxy,
e.g. 6.7~Gyr for a z=0.7 galaxy.}
\tablenotetext{b}{EXP,CSP: The latest formation redshift available
corresponds to the observed redshift of the galaxy {\sl minus}
0.1~Gyr, e.g. z$_{\rm F}>0.72$ for a z=0.7 galaxy.}
\end{deluxetable}
%%%%%%%%%%%%%%%%%%%%%%%%%%%%%%%%%%%%%%%

%%%%%%%%%%%%%%%%%%%%%%%%%%%%%%%%%%%%%%%
\section{Modelling the star formation histories of galaxies}

Our slitless spectra have good enough S/N down to the $i_{\rm
F775W}=24$~mag (AB) limit of the original sample. The S/N ranges from 30
down to 4--5 per pixel at the limiting magnitude. One of the main
advantages of G800L grism data for the analysis presented here is the
superb flux calibration of the spectra enabled by the optimal flat
fielding of the images.  We estimate flux calibration systematic
errors below 5\% within the spectral range of interest\citep{nor04}.

The wavelength coverage of the ACS/G800L grism (6000--9500\AA ) is
ideally suited for the analysis of old stellar populations at
z=0.4--1.5, because the prominent 4000\AA\ break falls within its
sensitivity range. Furthermore, the compact nature of early-type
galaxies and their homogeneous stellar populations minimises the
contamination and loss of spectral resolution compared to slitless
grism data of late-type or irregular galaxies.

\begin{figure}
\epsscale{1.2}
\plotone{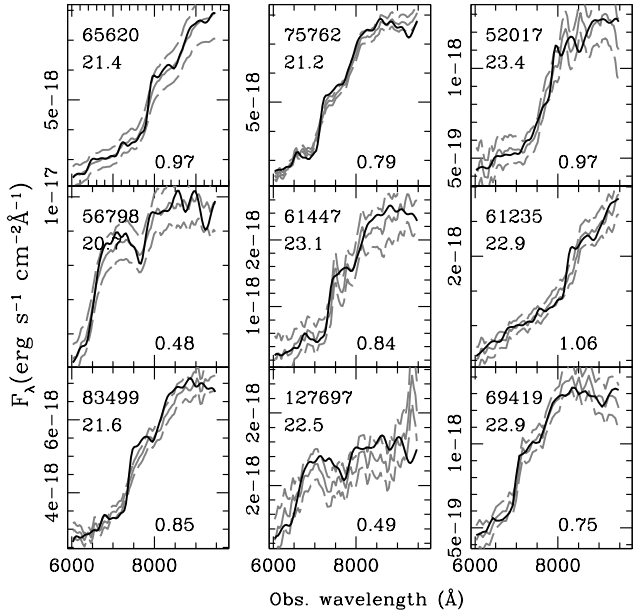}
\caption{A sample of best fits to the grism data of 'red' early-type galaxies
using models with chemical enrichment (CSP). Fits to the other two models
considered in this paper (EXP and 2BST) do not differ much from these. The
observed SEDs are shown in grey, including a 1$\sigma$ error envelopes (dashed lines).
The model fits are shown as black solid lines. Each galaxy is labelled
(from top to bottom) by their PEARS ID, $i_{\rm F775W}$ magnitude and redshift.
\label{fig:SEDsR}}
\end{figure}

%%%%%%%%%%%%%%%%%%%%%%%%%%%%%%%%%%%%%%%
\subsection{Methodology}

We follow a standard likelihood method fitting the low-resolution
spectra to grids of models. A proper extraction of the age and
metallicity distribution requires the use of composite models of
stellar populations. Simple stellar populations
(i.e. models with a single age and metallicity) will give
luminosity-weighted age estimates that can differ significantly from a
more physical mass-weighted age according to a composite stellar
population \citep{fy04,ap05,st07,ben08}. In order to assess systematic
differences related to the parameterisation of the star formation histories,
we consider the following four models (whose parameters are 
summarised in table~\ref{tab:params}).

\begin{itemize}
{
\item[I.] Simple Stellar Populations (SSP): SSPs are the building
  blocks of any population synthesis model. They correspond to a
  stellar population with a single age and metallicity. Although SSPs
  are often a fair approximation of a globular cluster, the longer
  timescales and more complex chemical enrichment history of a galaxy
  makes an SSP a very rough approximation in this case. Furthermore, the fact that
  a single age and metallicity are used to fit the data can introduce
  a significant difference between the SSP-determine age and a more
  physical mass-weighted age. Nevertheless, we include these models in
  the analysis for comparison, given that they are the simplest models
  from which the composite models are generated. This case explores a
  grid of $128\times 128$ SSPs over a wide range of ages and
  metallicities, as given in table~\ref{tab:params}.}  
\item[II.]{$\tau$-model (EXP): The SFH is described by an exponentially
  decaying function, with a timescale ($\tau$) and a formation
  redshift (z$_{\rm F}$), corresponding to the epoch at which star
  formation starts. The metallicity is the third free parameter and it
  is kept constant throughout each SFH.}
\item[III.]{2-Burst (2BST): Superposition of two simple stellar populations,
  described by four parameters: the age of the old ($t_{\rm O}$) and
  the young components ($t_{\rm Y}$), the mass fraction in young stars
  ($f_{\rm Y}$) and the metallicity of the system (assumed to be the
  same for both populations).}
\item[IV.]{Chemical Enrichment Model (CSP): We follow a consistent
  chemical enrichment code that incorporates the evolution of
  metallicity with the star formation rate.  The model is defined in
  \citet{fs00} and was previously applied to similar data from the
  GRAPES sample of early-types and bulges in the HUDF
  \citep{ap06b,nim09}. The formation history is described by four
  parameters: a star formation efficiency ($\nu$) controlling the
  transformation of gas into stars via a Schmidt law; an outflow
  parameter ($\beta$) which is the mass fraction of gas ejected as a
  result of the ongoing star formation rate; the exponential timescale
  of gas infall ($\tau_f$); and the formation epoch given by the
  redshift (z$_{\rm F}$) at which infall starts.}
\end{itemize}

\begin{figure}
\epsscale{1.2}
\plotone{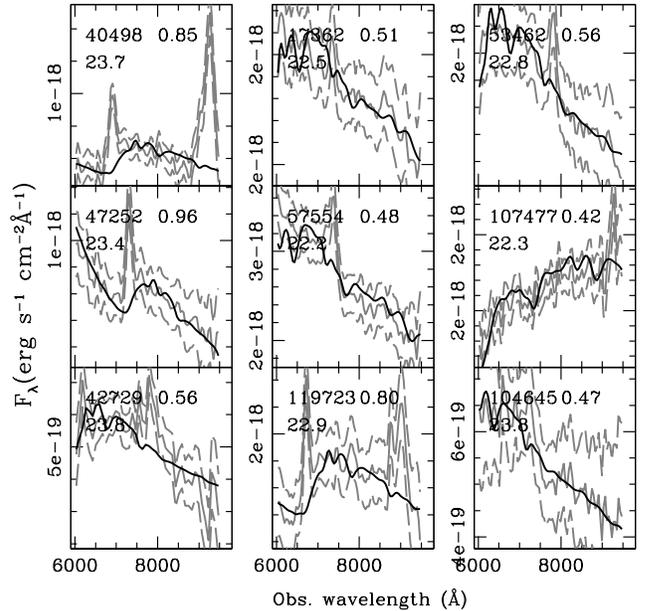}
\caption{Same as figure~\ref{fig:SEDsR} for a sample of 'blue'
early-type galaxies. The fits also correspond to CSP models. In this
case the fitting method excludes spectral windows of size $\pm
100$\AA\ around [\ion{O}{3}], [\ion{O}{2}], H$\beta$, and H$\alpha$
lines whenever the redshift moves them into the spectral range of
the grism data. (See \citet{str08} for details of emission line
galaxies in the PEARS HUDF spectra).
\label{fig:SEDsB}}
\end{figure}

\noindent
For a given choice of parameters, the models determine the
distribution of stellar ages and metallicities.  These distributions
are used to combine simple stellar populations from the latest 2007
models of Charlot \& Bruzual \citep[hereafter CB07,][ latest models in
preparation]{bc03,bruz07}, assuming a Chabrier initial mass function
\citep{chab03}. Notice the CB07 models have been recently updated with
respect to the high resolution spectra. However, at the low resolution
grism spectra considered in this paper, the difference between CB07
and the expected CB09 models is negligible (Charlot, priv. comm.).
The resulting spectra are degraded to the slitless PEARS SED, and
compared via a standard maximum likelihood method. Each model is
explored on a grid comprising between 32,000 and 65,000 realizations
for a range of parameters as described in table~\ref{tab:params}. The
best fit model is used to determine the age and metallicity of the
underlying populations, and the M/L required to translate flux into
stellar mass.

%%%%%%%%%%%%%%%%%%%%%%%%%%%%%%%%%%%%%%%
\subsection{Extracting star formation histories}

Figures~\ref{fig:SEDsR} and \ref{fig:SEDsB} show some of the best fits
of the CSP models for the `red' and `blue' galaxies, respectively.
The model SED is shown as a solid black line, and the observed SED is
shown in grey (including the 1$\sigma$ uncertainty as dashed
lines). Notice `blue' galaxies have emission lines, most notably
[\ion{O}{2}], [\ion{O}{3}] and H$\beta$. These lines are masked out in
the likelihood analysis. The blue spectra show a significant range of
ages, from large 4000\AA\ breaks (as in PID 119723) to very young
stars (e.g. 47252). In this paper the analysis only deals with stellar
populations, hence the presence of an AGN could significantly alter
the estimated ages of the `blue' galaxies. Nevertheless,
figure~\ref{fig:Phys1} shows that the subsample of `blue' galaxies is
dominated by low-mass systems, for which one would expect a
significant contribution to blue light from young stellar
populations. One could argue that these low stellar mass estimates are
biased because of the contribution from the AGN. However, the apparent
magnitude of all 'blue' galaxies in our sample is consistently fainter
than 'red' galaxies (see figure~\ref{fig:Obs}).
The `red' subsample (figure~\ref{fig:SEDsR}) looks much
more homogeneous, with a prominent 4000\AA break and no emission
lines. The SEDs also display the characteristic dip around the G-band
(rest-frame 4300\AA ) and for the lower redshift galaxies one can also
discern the Mg feature at rest-frame 5170\AA (see e.g. PID 122743 or 127697 in
figure~\ref{fig:SEDsR}).  Needless to say, the low resolution of the
G800L grism prevents us from doing any analysis 
of the absorption lines. However, the ``smoothed-out'' features of these
spectra can contribute to constraining the age and metallicity
distribution. Indeed, higher resolution spectra (R$\sim$2000) of Virgo
elliptical galaxies at very high signal-to-noise ratio (S/N$\gtrsim$
100\AA$^{-1}$) has shown that independent fits to the equivalent
widths or a direct fit to the SED give age distributions that are in good
agreement \citep{ben08}.

%%%%%%%%%%%%%%%%%%%%%%%%%%%%%%%%%%%%%%%%%%%%%
%%%%%%%%%%%%%%%%%%%%%%%%%%%%%%%%%%%%%%%%%%%%%
%%%%%%%%%%%%%    TABLE 3    %%%%%%%%%%%%%%%%%
%%%%%%%%%%%%%%%%%%%%%%%%%%%%%%%%%%%%%%%%%%%%%
%%%%%%%%%%%%%%%%%%%%%%%%%%%%%%%%%%%%%%%%%%%%%

\begin{deluxetable*}{ccrrrrrcrrr}
\tabletypesize{\scriptsize}
%\rotate
\tablecaption{Catalog of PEARS early-type galaxies: Physical properties (CSP models)
\label{tab:catphys}}
\tablewidth{13.5cm}
\tablehead{
\colhead{PID} & \colhead{M$_{\rm s}$} & \colhead{$U-V$} & \colhead{M$_V$} &
\colhead{R$_e$} & \colhead{$\langle$t$\rangle$} & \colhead{$\sigma_t^a$} & 
\colhead{$\langle\log Z/Z_\odot\rangle$} & \colhead{$\beta$} & \colhead{z$_{\rm F}$} 
& \colhead{log $\tau$}\\
\colhead{} & \colhead{$\times 10^{10}$M$_\odot$} & \colhead{} & \colhead{} &
\colhead{kpc} & \colhead{Gyr} & \colhead{Gyr} & \colhead{} & \colhead{} & \colhead{}
& \colhead{Gyr}}
\startdata
 65620 & 11.51 &  1.06  & $-$23.20 &  4.13 & 3.82 & 0.81 & $-$0.18 & 0.28 & 7.5 & $-$0.48\\
 56798 & 11.02 &  1.28  & $-$21.17 &  1.98 & 6.21 & 1.12 & $-$0.10 & 0.21 & 9.8 & $-$0.31\\
 83499 & 11.24 &  1.05  & $-$22.34 &  3.06 & 4.40 & 0.81 & $-$0.28 & 0.37 & 7.5 & $-$0.56\\
 75762 & 11.26 &  1.24  & $-$22.43 &  3.56 & 4.73 & 0.92 & $-$0.15 & 0.24 & 9.9 & $-$0.35\\
 61447 & 10.76 &  1.14  & $-$20.83 &  0.95 & 4.39 & 0.86 & $-$0.20 & 0.32 & 7.6 & $-$0.47\\
127697 & 10.29 &  1.17  & $-$19.39 &  1.31 & 5.81 & 1.06 & $-$0.11 & 0.17 & 5.5 & $-$0.53\\
 52017 & 10.19 &  0.80  & $-$20.95 &  0.90 & 2.71 & 0.71 & $-$0.30 & 0.41 & 3.2 & $-$0.50\\
 61235 & 10.86 &  1.06  & $-$22.16 &  1.55 & 3.64 & 0.77 & $-$0.22 & 0.33 & 9.8 & $-$0.48\\
 69419 & 10.35 &  1.08  & $-$20.42 &  1.26 & 4.51 & 0.89 & $-$0.18 & 0.26 & 6.0 & $-$0.48\\
 40498 &  9.33 &  0.02  & $-$19.47 &  1.79 & 1.05 & 0.53 & $-$0.36 & 0.45 & 0.8 & $-$0.07\\
 47252 &  8.69 & $-$0.49  & $-$19.87 &  0.86 & 0.10 & 0.04 & $-$0.94 & 0.41 & 0.9 & $-$0.26\\
 42729 &  8.83 &  0.37  & $-$18.32 &  1.06 & 0.66 & 0.38 & $-$0.38 & 0.45 & 0.7 & $-$0.12\\
 17362 &  9.72 &  0.47  & $-$19.45 &  1.24 & 2.35 & 0.74 & $-$0.47 & 0.58 & 1.2 & $-$0.22\\
 57554 &  9.85 &  0.48  & $-$19.58 &  1.04 & 1.98 & 0.71 & $-$0.54 & 0.61 & 1.0 & $-$0.19\\
119723 &  9.65 &  0.09  & $-$20.15 &  0.93 & 0.65 & 0.45 & $-$0.13 & 0.20 & 1.1 &  0.10\\
 53462 &  9.45 &  0.26  & $-$19.36 &  1.13 & 0.92 & 0.50 & $-$0.40 & 0.48 & 1.0 & $-$0.04\\
107477 & 10.31 &  1.41  & $-$19.16 &  2.21 & 2.27 & 0.75 & $-$0.36 & 0.46 & 1.0 & $-$0.22\\
104645 &  8.79 &  0.38  & $-$17.93 &  1.00 & 0.95 & 0.51 & $-$0.35 & 0.44 & 0.6 & $-$0.07\\
\enddata
\tablenotetext{a}{$\sigma_t$ is the root mean square of the
stellar age distribution.}
\tablecomments{A full version of table \ref{tab:catphys} is published in its 
entirety in the electronic edition of the {\it Astrophysical Journal}. 
A portion is shown here for guidance regarding its form and content.}
\end{deluxetable*}
%%%%%%%%%%%%%%%%%%%%%%%%%%%%%%%%%%%%%%%

\begin{figure}
\epsscale{1.2}
\plotone{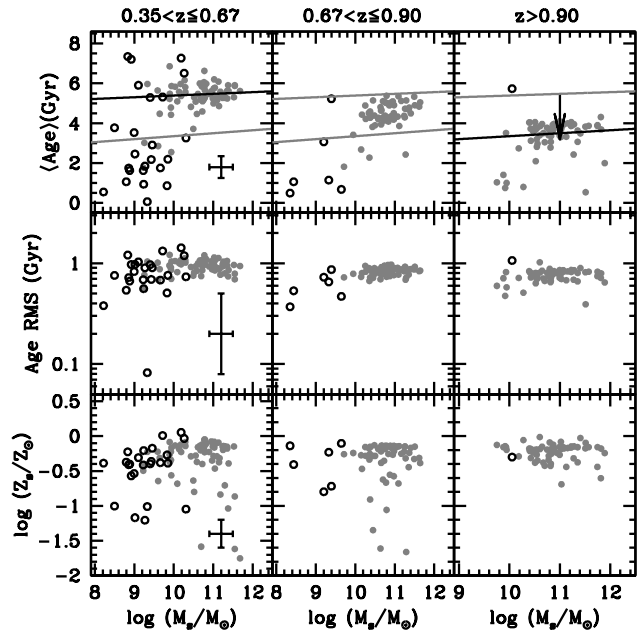}
\caption{Best fit average age (top), age spread (middle, defined as
the RMS of the distribution) and metallicity (bottom) for CSP
models. The lines in the top panel are simple least square fits to the
red galaxies for the lowest and highest redshift bins. These lines are
shown in black for the subsamples where the fit was done. The grey
lines are copies of these two fits, to guide the eye. The arrow in the
top right panel corresponds to the lookback time between the median
redshift of the high- and low redshift bins. On the left panels,
characteristic error bars are shown. 'Red' ('blue') early-type 
galaxies are shown as grey filled (black open) circles.
\label{fig:tZ}}
\end{figure}

\section{Results and Conclusions}

Figure~\ref{fig:tZ} shows the (mass-weighted) average age (top), age
scatter (defined as an RMS, middle) and metallicity (bottom) as a
function of stellar mass for three bins in redshift (top panel). These
estimates correspond to the chemical enrichment models (CSP). The best fit
star formation history from the models is used to compute the average and
RMS of the age distribution. 'Red'
and 'blue' early-types are shown as filled grey or open black circles,
respectively. Characteristic error bars are included. The figure
includes simple fits to the age distribution for the lowest redshift
sample (black line in the top-left panel) and the highest redshift bin
(black line in the top-right panel). The lines in grey are just copies
of these two lines in the other panels, to guide the eye. The arrow in
the top-right panel is the expected age difference between the high
and the low redshift bin (i.e. pure passive evolution).

The figure allows us to understand the blue cloud-red sequence diagram
of the sample. Our blue-cloud spheroidal galaxies are younger systems
with roughly similar metallicities to the red sequence
distribution. It is worth mentioning that the CSP models suggest all
our galaxies have a similar {\sl spread} of stellar ages, and it is
only the {\sl average} age that correlates with stellar mass. This
result is consistent with the more detailed analysis of \citet{ben08}
on high-S/N spectra of elliptical galaxies in the Virgo cluster. The
lines shown in the top panel illustrate that passive evolution over
this redshift range can explain the 'red' spheroidals -- which
represent the large majority of the sample. The average ages of
the subsample at lower redshift ($0.35\leq$z$\leq 0.67$; {\sl
top-left}) suggest that the correlation between stellar age and mass
is such that instead of a monotonic trend, the scatter in age
increases in galaxies with stellar masses below 10$^{10}$M~$_\odot$
\citep[see e.g.][]{cald03,gal06,yam06,ben08}.

\begin{figure}
\epsscale{1.2}
\plotone{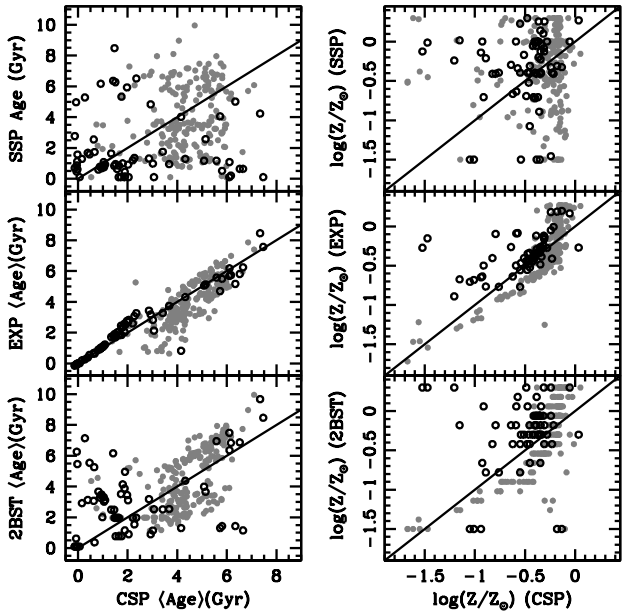}
\caption{
Comparison between the predicted average ages and metallicities for the
models explored in this paper (see text for details). The solid line
gives the 1:1 mapping. Solid (open) circles correspond to 'red' ('blue')
early-type galaxies.
\label{fig:tZcomp}}
\end{figure}

In order to check systematic effects arising from the specific way to
parameterise the models, we compare in figure~\ref{fig:tZcomp} the
average age and metallicity distribution of the three models
considered. There is overall good agreement between the CSP and the
EXP models (top panels), whereas the age and metallicities of SSP and
2BST models differ significantly with respect to CSP models. 
Notice that best fit SSP metallicities are often very low, as the best fit
wanders along the age-metallicity degeneracy, affecting the estimated
ages. Composite models have been found to give more realistic
metallicity constraints than SSPs \citep{fy04}.
However, we cannot look into the reduced $\chi^2$ values to rule out
one model against another: all models give equally acceptable values
of $\chi_r^2\sim 1$. Observations at higher spectral resolution dot
not fare better at disentangling this degeneracy \citep{ben08}.

It is also worth mentioning that for the composite
models (2BST, EXP and CSP) there is good agreement between the
best-fit metallicities for the red galaxies -- within a
realistic $\sim 0.3$~dex uncertainty. Blue galaxies, as expected,
give more discrepant results, mainly caused by the lack of
spectral features in the low-resolution PEARS SEDs which makes
metallicity estimates more uncertain.

\begin{figure}
\epsscale{1.2}
\plotone{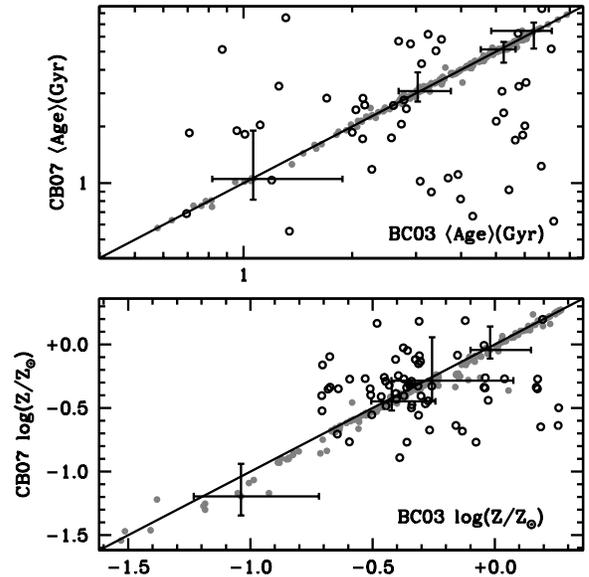}
\caption{We illustrate here the effect of a different choice
of stellar population models. 
The average age ({\sl top}) and metallicity ({\sl bottom}) of our
models, based on the latest Charlot \& Bruzual (2007) models, is
compared with the predictions when the previous version of 
population synthesis models are used  \citep{bc03}. 
Notice `red' galaxies (filled circles) give similar
results, whereas the younger populations found in the `blue' subsample
(open circles) depend strongly on the choice of population models.
\label{fig:bc03}}
\end{figure}

Another check of our results was done with respect to the older
version of the population synthesis models of \citet{bc03}. In
figure~\ref{fig:bc03} we compare the average age and metallicity when
using either the 2003 models (horizontal axes) or the improved CB07
models (vertical axes). The discrepancy between both versions is most
prominent for the 'blue' subsample (open circles), which reflects
the different treatment of some of the young-intermediate age phases
of stellar evolution. The difference of
age or metallicity between BC03 and CB07 models for older populations
('red' galaxies, grey circles) is negligible within the error bars.

\begin{figure}
\epsscale{1.2}
\plotone{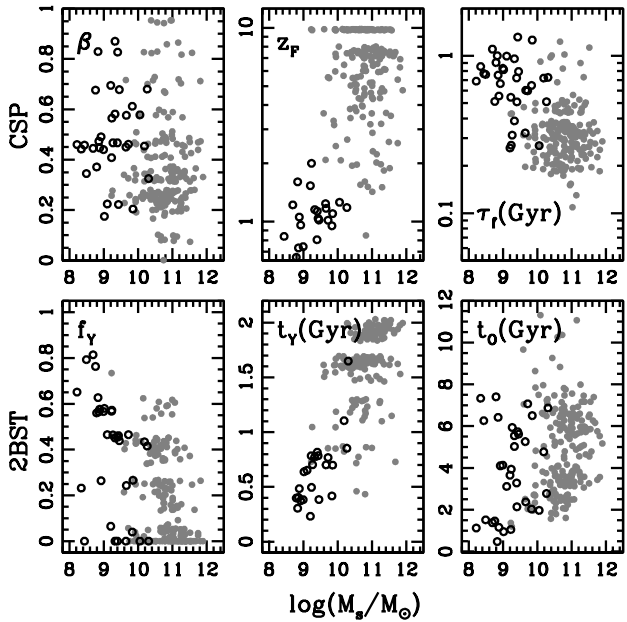}
\caption{Best-fit parameters for a continuous model including
chemical enrichment (CSP, top) or a two-burst scenario (2BST,
bottom). The CSP parameters are (from left to right), the outflow
fraction ($\beta$), formation epoch (z$_{\rm F}$; defined as the time
when gas infall starts) and the exponential timescale of gas infall
($\tau_f$). The 2BST parameters are (also from left to right) the mass
fraction in young stars ($f_{\rm Y}$); young stellar age ($t_{\rm Y}$) 
; and old stellar age ($t_{\rm O}$).
\label{fig:params}}
\end{figure}

The average and spread of the age distribution shown in
figure~\ref{fig:tZ} can be shown with respect to the phenomenological
parameters used to describe the star formation
history. Figure~\ref{fig:params} shows the best-fit parameters for the
CSP (top) and the 2BST (bottom) models as a function of stellar
mass. A large spread is found for the fraction
of gas and metals ejected in outflows ($\beta$), whereas the gas
infall timescale is overall rather short ($\tau_f\lesssim
1$~Gyr). Over the 0.4$<$z$<$1.2 range probed by our sample --
corresponding to a cosmic age between 5 and 10 Gyr (with 13.5~Gyr
being our present age) -- it is the formation epoch (top middle panel)
what mainly correlates with stellar mass to give the age-mass trend
shown in figure~\ref{fig:tZ}. Our 'blue' early-types have formation
epochs z$_{\rm F}\sim 1$, which -- given the redshift of the sample --
imply a large amount of young stars.

If we consider instead a two-burst model (2BST; bottom panels), the
main driver of the age-mass relation is given by the young stellar
mass fraction ($t_{\rm Y}$), although the mass fraction in young stars
($f_{\rm Y}$) also shows a correlation with mass. 'Blue' early-types
have high values of $f_{\rm Y}$. In general, we find that significant
masses in young stars (open circles) only appear in early-type
galaxies with masses $\lesssim 10^{10}$M$_\odot$. However, notice
there is a fraction of 'red' early-type galaxies -- which
live on the red sequence (figure~\ref{fig:CMR}) -- with a significant
amount of young stars ($f_{\rm Y}\gtrsim 0.4$). This is consistent
with the observed increase of the recent star formation in spheroidal
galaxies with redshift \citep{kav08}.

Both a smooth description of the star formation history (CSP) and a
more abrupt model (2BST) confirm that red-sequence spheroidal galaxies
with masses above $\sim 10^{11}$M$_\odot$ have the bulk of their
stellar populations formed at z$_{\rm F}\gtrsim 3$. This result
confirms previous findings \citep[see e.g.][]{tan96,vaz97,perez08,wik08,man09}
although our work improves on the analysis, since we use SEDs instead of
photometry to constrain the age distribution, and we consider a large
volume of parameter space to describe the star formation history.

\begin{figure*}
\epsscale{1.1}
\plotone{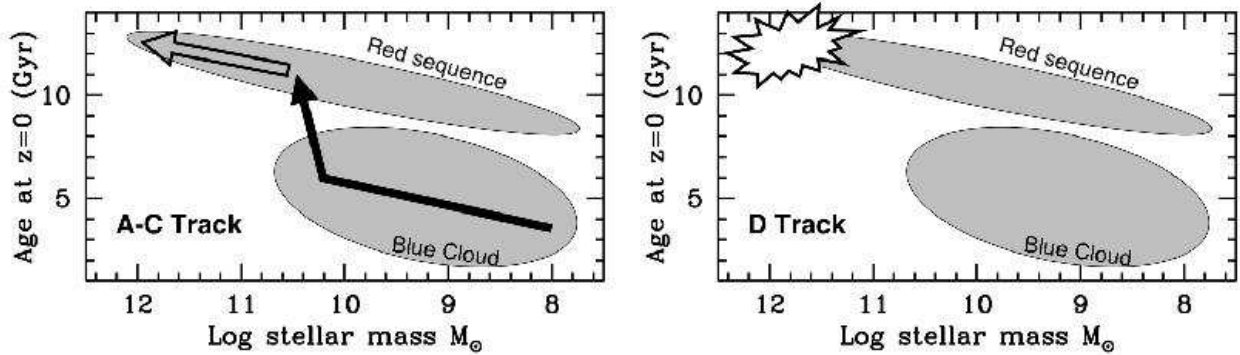}
\caption{Our proposed additional channel for the formation of galaxies
on the red sequence is shown as 'D track' (right), in comparison with
the standard picture from \citet{fab07} (A-C track, left). The age
distribution of our sample of moderate redshift early-type galaxies
along with recent findings of a lack of evolution of the comoving
number density or the intrinsic colour distribution of massive
galaxies \citep[see e.g.][]{fon06,con07,fl09b} suggest an {\sl in
situ} build-up of the stellar populations bypassing any trajectory in
the blue cloud (except for the very short-lived formation stage
corresponding to an intense, highly efficient process of star
formation). This proposed channel can be justified by the cold
accretion streams found in recent numerical simulations \citep{dek09}.
\label{fig:cartoon}}
\end{figure*}

The old populations found in massive early-type galaxies in
combination with the lack of number density evolution \citep[see
e.g.][]{fl09b} confirms that new channels must be involved to explain
the presence of galaxies on the red sequence at the massive end. In
analogy to the cartoon version of the galaxy evolution tracks A-C in
figure~10 of \citet{fab07}, we propose a new one (track D) in the right
panel of our figure~\ref{fig:cartoon}. Tracks A-C (left panel) follow
the standard growth via wet mergers on the blue cloud, followed by
quenching of star formation (the near vertical black arrow) after
which growth to the top of the red sequence consists of dry mergers
(white arrow). This option is still controversial, since different
semi-analytic models predict a very different redshift evolution of
the number density \citep[][Hopkins \etal, in preparation]{fl09b}. On
the other hand, the recent work of \citet{dek09} based on the
Mare~Nostrum cosmological simulations suggest a significant amount of
massive galaxy growth via cold accretion at the intersection of cosmic
filaments. This process would bypass the evolution in the blue cloud,
as visualised by our track D on the right panel of
figure~\ref{fig:cartoon}. In this case, most of the stellar
component in a massive galaxy would grow {\sl in situ}, over a 
short and early period, explaining
the old populations found in massive galaxies (i.e. the ``explosion'' 
symbol in our track D). Furthermore, this
process would be very efficient, turning gas into stars over short
timescales, thereby explaining the small infall timescales found
(figure~\ref{fig:params}). The fact that formation epoch correlates
with stellar mass would indicate the connection between the stellar
mass of the galaxy and the significance of the overdensity from which
these galaxies grow. Lower masses correspond to 'lower-$\sigma$'
fluctuactions thereby collapsing later. Nevertheless, it is worth
emphasizing that the channel proposed by \citet{dek09} would dominate
at the most massive end. Intermediate mass galaxies will have a more
complex history, with both tracks A-C and D contributing to their
formation. The picture is unfortunately not fully explained, as
massive galaxies also experience a very significant size evolution
between z$\sim$2 and z=0 \citep{daddi05,vdk08,bui08}, an issue not explained
yet in the context of \citet{dek09}.

As a final note, we may speculate on the importance of the formation
redshifts shown in the top-middle panel of
figure~\ref{fig:params}. The best-fit values range between z$_{\rm
F}\simeq$0.5 to z$_{\rm F}\sim$10, as expected in hierarchical
formation scenarios that produce a broad peak in the cosmic SFH at
z$\sim$1--2. However, we also note that the highest formation
redshifts pile up either between z$_{\rm F}\simeq$7--8 or around
z$_{\rm F}\simeq$10. While the estimates of the current
spectrophotometry and SED modelling is undoubtedly not accurate enough
to state the difference between the epochs z$_{\rm F}\simeq$7--8 and
z$_{\rm F}\simeq$10 with any great confidence, it is tempting to
notice that these redshifts coincide remarkably well with the epochs
in WMAP-year5 cosmology where the peak of the Pop III star-formation
should have occurred at z$\simeq$10.8$\pm$1.8 in order to produce the
WMAP-year5 polarization optical depth \citep{jo09}, and with the epoch
of z$_{\rm F}\simeq$7--8 in which the global component of Pop II stars
should have started shining in dwarf galaxies in order to finish
reionization by z$\simeq$6. Future data and modeling should
concentrate on whether these epochs can be defined any more accurately
than possible here. This analysis may be feasible with the upcoming
HST/WFC3 grism spectra of the earliest type galaxies at 1$\le$z$\le$3
in the near future.

\acknowledgments
We would like to thank St\'ephane Charlot for making available the
latest Charlot \& Bruzual population synthesis models. The anonymous
referee is gratefully acknowledged for useful comments. The project
presented in this paper made use of the UCL High Performance Computing
facility Legion. This research was supported in part by grants
HST-GO-10530 and HST-GO-9793 from STScI, which is operated by AURA for
NASA under contract NAS 5-26555.

Facilities: \facility{\hst\ (ACS)}

\bibliographystyle{aj.bst}
\bibliography{allrefs}

%%%%%%%%%%%%%%%%%%%%%%%%%%%%%%%%%%%%%%%
%%%%%%%%    REFERENCES
%%%%%%%%%%%%%%%%%%%%%%%%%%%%%%%%%%%%%%%

\begin{thebibliography}{}
\bibitem[\protect\citeauthoryear{Alexander \etal}{2003}]{alex03}
Alexander, D.~M., \etal 2003, AJ, 126, 539

\bibitem[\protect\citeauthoryear{Arag\'on-Salamanca \etal}{2003}]{as93}
  Arag\'on-Salamanca, A., Ellis, R.~S., Couch, W.~J. \& 
  Carter, D. 1993, \mnras, 262, 764

\bibitem[\protect\citeauthoryear{Beckwith \etal}{2006}]{sb06}
  Beckwith, S.~V.~W., \etal 2006, \aj, 132, 1729

\bibitem[\protect\citeauthoryear{Bell \etal}{2004}]{Bell04}
  Bell, E.~F. \etal 2004, \apj, 608, 752

\bibitem[\protect\citeauthoryear{Bershady \etal}{2000}]{Ber00}
  Bershady, M.~A., Jangren, A. \& Conselice, C.~J. 2000, \aj, 119, 2645
  
\bibitem[\protect\citeauthoryear{Bernardi \etal}{2003}]{ber03}
  Bernardi, M., \etal  2003, \aj, 125, 1882

\bibitem[\protect\citeauthoryear{Blakeslee \etal}{2003}]{blk03}
  Blakeslee, J.~P., \etal 2003, \apj, 596, 143 
  
\bibitem[\protect\citeauthoryear{Blanton \etal}{2003}]{blan03}
  Blanton, M.~R., \etal 2003, \apj, 594, 186

\bibitem[\protect\citeauthoryear{Blanton \etal}{2005}]{blan05}
  Blanton, M.~R., \etal 2005, \aj, 129, 2562

\bibitem[\protect\citeauthoryear{Borch \etal}{2006}]{borch06}
  Borch, A., \etal 2006, \aap, 453, 869

\bibitem[\protect\citeauthoryear{Bower \etal}{1992}]{ble92}
  Bower, R.~G., Lucey, J.~R. \& Ellis, R.~S. 1992, \mnras, 254, 601
  
\bibitem[\protect\citeauthoryear{Brammer \& van~Dokkum}{2007}]{bram07}
  Brammer, G.~B. \& van Dokkum, P.~G. 2007, \apj, 654, 107	

\bibitem[\protect\citeauthoryear{Bruzual \& Charlot}{2003}]{bc03}
  Bruzual, G. \& Charlot, S. 2003, \mnras, 344, 1000

\bibitem[\protect\citeauthoryear{Bruzual}{2007}]{bruz07}
  Bruzual, G. 2007, IAU No. 241 Symp. Procs.
  "Stellar populations as building blocks of galaxies", 
  eds. A. Vazdekis and R.~F. Peletier, Cambridge, arXiv:astro-ph/0703052

\bibitem[\protect\citeauthoryear{Buitrago \etal}{2008}]{bui08}
  Buitrago, F., Trujillo, I., Conselice, C., Bouwens, R.~J., 
  Dickinson, M. \& Yan, H., 2008, ApJ, 687, L61
  
\bibitem[\protect\citeauthoryear{Caldwell \etal}{2003}]{cald03}
  Caldwell, N., Rose, J.~A. \& Concannon, K.~D. 2003, \aj, 125, 2891

\bibitem[\protect\citeauthoryear{Cappellari \etal}{2009}]{cap09}
  Cappellari, M., \etal, 2009, arXiv:0906.3648

\bibitem[\protect\citeauthoryear{Chabrier}{2003}]{chab03}
  Chabrier, G. 2003, \pasp, 15, 763

\bibitem[\protect\citeauthoryear{Cenarro \& Trujillo}{2009}]{ct09}
  Cenarro, J. \& Trujillo, I. 2009, \apj, 696, L43

\bibitem[\protect\citeauthoryear{Cimatti \etal}{2004}]{cim04}
  Cimatti, A., \etal 2004, \nat, 430, 184

\bibitem[\protect\citeauthoryear{Cimatti \etal}{2008}]{cim08}
  Cimatti, A., \etal 2008, A\&A, 482, 21

\bibitem[\protect\citeauthoryear{Conselice}{2003}]{con03}
  Conselice, C.~J. 2003, \apjs, 147, 1

\bibitem[\protect\citeauthoryear{Conselice \etal}{2007}]{con07}
  Conselice, C.~J., \etal 2007, \mnras, 381, 962
  
\bibitem[\protect\citeauthoryear{Cool \etal}{2008}]{cool08}
  Cool, R~J., \etal 2008, \apj, 682, 919

\bibitem[\protect\citeauthoryear{Croton \etal}{2005}]{crot05}
  Croton, D.~J., \etal 2005, \mnras, 356, 1155

\bibitem[\protect\citeauthoryear{Croton \etal}{2006}]{crot06}
  Croton, D.~J., \etal 2006, \mnras, 365, 11

\bibitem[\protect\citeauthoryear{Daddi \etal}{2005}]{daddi05}
  Daddi, E., \etal 2005, \apj, 626, 680

\bibitem[\protect\citeauthoryear{Dekel \etal}{2009}]{dek09}
  Dekel, A., \etal 2009, \nat, 457, 451

\bibitem[\protect\citeauthoryear{De~Lucia \etal}{2006}]{Delu06}
  De Lucia, G., Springel, V., White, S.~D.~M., Croton, D. \&
  Kauffmann, G. 2006, \mnras,  366, 499

\bibitem[\protect\citeauthoryear{Dressler \etal}{1990}]{dres90}
  Dressler, A., Gunn, J.E., 1990, ASPC, 10, 200

\bibitem[\protect\citeauthoryear{Driver \etal}{1998}]{driv98}
  Driver, S.~P., Fern\'andez-Soto, A., Couch, W.~J.,
  Odewahn, S.~C., Windhorst, R.~A., Phillips, S., Lanzetta, K. 
  \& Yahil, A. 1998, \apj, 496, L93

\bibitem[\protect\citeauthoryear{Dunkley \etal}{2009}]{jo09}
  Dunkley, J., \etal 2009, \apjs, 180, 306 

\bibitem[\protect\citeauthoryear{Dunlop \etal}{1996}]{dun96}
  Dunlop, J., Peacock, J., Spinrad, H., Dey, A., Jimenez, R., 
  Stern, D. \& Windhorst, R. 1996, \nat, 381, 581
  
\bibitem[\protect\citeauthoryear{Eggen \etal}{1962}]{els62}
  Eggen, O.~J., Lynden-Bell, D. \& Sandage, A.~R. 1962, \apj, 136, 748

\bibitem[\protect\citeauthoryear{Ellis \etal}{1997}]{rse97}
  Ellis, R.~S., Smail, I., Dressler, A., Couch, W.~J., Oemler, A.~Jr., 
  Butcher, H. \& Sharples, R.~M. 1997, \apj, 483, 582
 
\bibitem[\protect\citeauthoryear{Faber \etal}{2007}]{fab07}
  Faber, S.~M., \etal 2007, \aj, 665, 265

\bibitem[\protect\citeauthoryear{Fan \etal}{2008}]{fan08}
  Fan, L., Lapi, A., De~Zotti, G. \& Danese, L. 2008, \apj, 689, L101

\bibitem[\protect\citeauthoryear{Ferreras \etal}{1999}]{fcs99}
  Ferreras, I., Charlot, S. \& Silk, J. 1999, \apj, 521, 81
 
\bibitem[\protect\citeauthoryear{Ferreras \& Silk}{2000}]{fs00}
  Ferreras, I. \& Silk, J. 2000, \apj, 532, 193

\bibitem[\protect\citeauthoryear{Ferreras \& Silk}{2003}]{fs03} 
  Ferreras, I. \& Silk, J. 2003, \mnras, 344, 455
  
\bibitem[\protect\citeauthoryear{Ferreras \& Yi}{2004}]{fy04} 
  Ferreras, I. \& Yi, S.~K. 2004, \mnras, 350, 1322

\bibitem[\protect\citeauthoryear{Ferreras \etal}{2005}]{fl05}
  Ferreras, I., Lisker, T., Carollo, C.~M., Lilly, S.~J. \& 
  Mobasher, B. 2005, \apj, 635, 243
  
\bibitem[\protect\citeauthoryear{Ferreras \etal}{2006}]{pca}
  Ferreras, I., Pasquali, A., de~Carvalho, R.~R., de~la~Rosa, I.~G. \&
  Lahav, O. 2006, \mnras, 370, 828
  
\bibitem[\protect\citeauthoryear{Ferreras \etal}{2009a}]{fl09a}
  Ferreras, I., Lisker, T., Pasquali, A. \& Kaviraj, S., 2009a, \mnras, 395, 554
  
\bibitem[\protect\citeauthoryear{Ferreras \etal}{2009b}]{fl09b}
  Ferreras, I., Lisker, T., Pasquali, A., Khochfar, S. \&
  Kaviraj, S. 2009b, \mnras, 396, 1573
  
\bibitem[\protect\citeauthoryear{Fontana \etal}{2006}]{fon06}
  Fontana, A., \etal 2006, \aap, 459, 745

\bibitem[\protect\citeauthoryear{Gallazzi \etal}{2006}]{gal06}
  Gallazzi, A., Charlot, S., Brichmann, J. \& White, S.~D.~M.
  2008, \mnras, 387, 1253

\bibitem[\protect\citeauthoryear{Gladders \etal}{1998}]{glad98}
  Gladders, M.~D., Lopez-Cruz, O., Yee, H.~K.~C. \& 
  Kodama, T. 1998, \apj, 501, 571
  
\bibitem[\protect\citeauthoryear{Glazebrook \etal}{2004}]{glaz04}
  Glazebrook, K., \etal 2004, \nat, 430, 181
  
\bibitem[\protect\citeauthoryear{Hathi \etal}{2009}]{nim09}
  Hathi, N.~P., Ferreras, I., Pasquali, A., Malhotra, S., 
  Rhoads, J.~E., Pirzkal, N., Windhorst, R.~A. \& Xu, C. 2009, 
  \apj, 690. 1866

\bibitem[\protect\citeauthoryear{Hopkins \etal}{2006}]{hop06}
  Hopkins, P.~F., Somerville, R.~S., Hernquist, L., Cox, T.~J., 
  Robertson, B. \& Li, Y. 2006, \apj, 652, 864

\bibitem[\protect\citeauthoryear{Hopkins \etal}{2008}]{hop08}
  Hopkins, P.~F., Cox, T.~J., Keres, D. \& Hernquist, L. 2008, \apjs, 175, 390

\bibitem[\protect\citeauthoryear{Hopkins \etal}{2009}]{hop09}
  Hopkins, P.~F., Hernquist, L., Cox, T.~J., Keres, D. \&
  Wuyts, S. 2009, \apj, 691, 1424

\bibitem[\protect\citeauthoryear{Hubble}{1926}]{edwin}
  Hubble, E.~D. 1926, \apj, 64, 321
  
\bibitem[\protect\citeauthoryear{Kaviraj \etal}{2005}]{kav05}
  Kaviraj, S., Devriendt, J.~E.~G., Ferreras, I. \& Yi, S.~K. 2005, \mnras, 360, 60
  
\bibitem[\protect\citeauthoryear{Kaviraj \etal}{2007}]{kav07}
  Kaviraj, S., \etal 2007, \apjs, 173, 619

\bibitem[\protect\citeauthoryear{Kaviraj \etal}{2008}]{kav08}
  Kaviraj, S., \etal 2008, \mnras, 388, 67 
  
\bibitem[\protect\citeauthoryear{Kelson \etal}{2000}]{kel00}
  Kelson, D.~D., Illingworth, G.~D., van~Dokkum, P.~G. \& 
  Franx, M. 2000, \apj, 531, 184

\bibitem[\protect\citeauthoryear{Kelson \etal}{2001}]{kel01}
  Kelson, D.~D., Illingworth, G.~D., Franx, M. \& 
  van~Dokkum, P.~G. 2001, \apj, 552, 17

\bibitem[\protect\citeauthoryear{Khochfar \& Silk}{2006a}]{ks06a}
  Khochfar, S. \& Silk, J. 2006a, \mnras, 370, 902

\bibitem[\protect\citeauthoryear{Khochfar \& Silk}{2006b}]{ks06b}
  Khochfar, S. \& Silk, J. 2006b, \apj, 648, 21 
  
\bibitem[\protect\citeauthoryear{Kodama \& Arimoto}{1997}]{ka97}
  Kodama, T. \& Arimoto, N. 1997, \aap, 320, 41

\bibitem[\protect\citeauthoryear{Kodama \etal}{1998}]{kod98}
  Kodama, T., Arimoto, N., Barger, A.~J. \& 
  Arag\'on-Salamanca, A. 1998, A\&A, 334, 99

\bibitem[\protect\citeauthoryear{Kodama \etal}{2007}]{kod07}
  Kodama, T., \etal 2007, \mnras, 377, 1717

\bibitem[\protect\citeauthoryear{Komatsu \etal}{2009}]{wmap5}
  Komatsu, E. \etal 2009, \apjs, 180, 330

\bibitem[\protect\citeauthoryear{Kriek \etal}{2008}]{kriek08}
  Kriek, M., van~der~Wel, A., van~Dokkum, P.~G., Franx, M. \&
  Illingworth, G.~D. 2008, \apj, 682, 896

\bibitem[\protect\citeauthoryear{Larson}{1974}]{lar74} 
  Larson, R.~B. 1974, \mnras, 166, 585

\bibitem[\protect\citeauthoryear{Lintott \etal}{2006}]{lfl06} 
  Lintott, C. J., Ferreras, I. \& Lahav, O. 2006, \apj, 648, 826
  
\bibitem[\protect\citeauthoryear{Lotz \etal}{2004}]{lotz04}
  Lotz, J.~M., Primack, J. \& Madau, P. 2004, \aj, 128, 163

\bibitem[\protect\citeauthoryear{Luo \etal}{2008}]{luo08}
Luo, B., \etal 2008, ApJS, 179, 19

\bibitem[\protect\citeauthoryear{Mancini \etal}{2009}]{man09}
 Mancini, C., Matute, I., Cimatti, A., Daddi, E., Dickinson, M., 
 Rodighiero, G., Bolzonella, M. \& Pozzetti, L., 2009, \aap, in press, arXiv:0901.3341

\bibitem[\protect\citeauthoryear{Maraston \etal}{2006}]{mar06}
  Maraston, C., \etal 2006, \apj, 652, 85
  
\bibitem[\protect\citeauthoryear{McCarthy \etal}{2004}]{McCa04}
  McCarthy, P.~J., \etal 2004, \apj, 614, 9
  
\bibitem[\protect\citeauthoryear{McIntosh \etal}{2005}]{McIn05}
  McIntosh, D.~H., \etal 2005, \apj, 632, 191

\bibitem[\protect\citeauthoryear{Mobasher \etal}{2004}]{mob04} 
  Mobasher, B., \etal 2004, \apj, 600, L167 

\bibitem[\protect\citeauthoryear{Monaco \etal}{2007}]{mon07}
  Monaco, P., Fontanot, F. \& Taffoni, G. 2007, \mnras, 375, 1189

\bibitem[\protect\citeauthoryear{Nelan \etal}{2005}]{nel05}
  Nelan, J.~E., \etal 2005, \apj, 632, 137
  
\bibitem[\protect\citeauthoryear{Nelson \etal}{2001}]{nel01}
  Nelson, A.~E., Gonz\'alez, A.~H., Zaritsky, D. \&
  Dalcanton, J.~J. 2001, \apj, 563, 629

\bibitem[\protect\citeauthoryear{Pasquali \etal}{2005}]{ap05}
  Pasquali, A., Larsen, S., Ferreras, I., Gnedin, O.~Y., Malhotra, S., 
  Rhoads, J.~E., Pirzkal, N. \& Walsh, J.~R. 2005, \aj, 129, 148

\bibitem[\protect\citeauthoryear{Pasquali \etal}{2006a}]{ap06a}
  Pasquali, A., Pirzkal, N., Larsen, S., Walsh, J.~R. \& 
  K\"ummel, M. 2006a, \pasp, 118, 270

\bibitem[\protect\citeauthoryear{Pasquali \etal}{2006b}]{ap06b}
  Pasquali, A., \etal  2006b, \apj, 636, 115

\bibitem[\protect\citeauthoryear{P\'erez-Gonz\'alez \etal}{2008}]{perez08}
  P\'erez-Gonz\'alez, P.~G., \etal 2008, \apj, 675, 234

\bibitem[\protect\citeauthoryear{Pirzkal \etal}{2004}]{nor04}
  Pirzkal, N., \etal 2004, \apjs, 154, 501

\bibitem[\protect\citeauthoryear{Popesso \etal}{2008}]{pop08}
  Popesso, P., \etal 2008, \aap, submitted, arXiv:0802.2930

\bibitem[\protect\citeauthoryear{Rakos \& Schombert}{1995}]{rs95}
  Rakos, K.~D. \& Schombert, J.~M., 1995, \apj, 439, 47
  
\bibitem[\protect\citeauthoryear{Rogers \etal}{2007}]{ben07}
  Rogers, B., Ferreras, I., Lahav, O., Bernardi, M., Kaviraj, S. \&
  Yi, S.~K. 2007, \mnras, 382, 750

\bibitem[\protect\citeauthoryear{Rogers \etal}{2008}]{ben08}
  Rogers, B., Ferreras, I., Peletier, R.~F. \&
  Silk, J. 2008, \mnras, submitted, arXiv:0812.2029

\bibitem[\protect\citeauthoryear{Saracco \etal}{2005}]{sar05}
  Saracco, P., \etal 2005, \mnras, 357, 40

\bibitem[\protect\citeauthoryear{Scannapieco \etal}{2005}]{evan05}
  Scannapieco, E., Silk, J. \& Bouwens, R. 2005, \apj, 635, L13

\bibitem[\protect\citeauthoryear{Scarlata \etal}{2007}]{sca07}
  Scarlata, C., \etal 2007, \apjs, 172, 494
  
\bibitem[\protect\citeauthoryear{Schawinski \etal}{2007}]{scha07}
  Schawinski, K., \etal 2007, \apjs, 173, 512

\bibitem[\protect\citeauthoryear{Serra \& Trager}{2007}]{st07}
  Serra, P. \& Trager, S.~C. 2007, \mnras, 374, 769

\bibitem[\protect\citeauthoryear{Shen \etal}{2003}]{shen03}
  Shen, S., Mo, H.~J., White, S.~D.~M., Blanton, M.~R., Kauffmann, G., Voges, W., 
  Brinkmann, J. \& Csabai, I. 2003, \mnras, 343, 978
  
\bibitem[\protect\citeauthoryear{Somerville \etal}{2008}]{som08}
  Somerville, R.~S., Hopkins, P.~F., Cox, T.~J., Robertson, B.~E. \&
  Hernquist, L. 2008, \mnras, 391, 481

\bibitem[\protect\citeauthoryear{Spinrad \etal}{1997}]{hy97}
  Spinrad, H., Dey, A., Stern, D., Dunlop, J., Peacock, J., 
  Jimenez, R. \& Windhorst, R. 1997, \apj, 484, 581

\bibitem[\protect\citeauthoryear{Stanford \etal}{1998}]{sed98}
  Stanford, S.~A., Eisenhardt, P.~R., Dickinson, M. 1998, \apj, 492, 461

\bibitem[\protect\citeauthoryear{Strateva \etal}{2001}]{strat01}
  Strateva, I., \etal 2001, \aj, 122, 1861
  
\bibitem[\protect\citeauthoryear{Straughn \etal}{2008}]{str08}
  Straughn, A. \etal 2008, \aj, 135, 1624

\bibitem[\protect\citeauthoryear{Tantalo \etal}{1996}]{tan96}
  Tantalo, R., Chiosi, C., Bressan, A. \& Fagotto, F. 1996, \aap, 311, 361

\bibitem[\protect\citeauthoryear{Thomas \etal}{1999}]{thom99}
  Thomas, D., Greggio, L. \& Bender, R. 1999, \mnras, 302, 537
  
\bibitem[\protect\citeauthoryear{Thomas \etal}{2005}]{thom05}
  Thomas, D., Maraston, C., Bender, R. \& Mendes de Oliveira, C.
  2005, \apj, 621, 673
  
\bibitem[\protect\citeauthoryear{Trager \etal}{2000}]{sct00}
  Trager, S.~C., Faber, S.~M., Worthey, G. \&
  Gonz\'alez, J.~J. 2000, \aj, 120, 165

\bibitem[\protect\citeauthoryear{Treu \etal}{2005a}]{Treu05a}
  Treu, T., Ellis, R.~S., Liao, T.~X. \& van Dokkum, P.~G. 2005a, \apj, 622, 5
  
\bibitem[\protect\citeauthoryear{Treu \etal}{2005b}]{Treu05b}
  Treu, T., \etal\ 2005b, \apj, 633, 174
  
\bibitem[\protect\citeauthoryear{Trujillo \etal}{2006}]{truj06}
  Trujillo, I., \etal 2006, \apj, 650, 18
  
\bibitem[\protect\citeauthoryear{Trujillo \etal}{2007}]{truj07}
  Trujillo, I., Conselice, C.~J., Bundy, K., Cooper, M.~C., 
  Eisenhardt, P. \& Ellis, R.~S. 2007, \mnras, 382, 109

\bibitem[\protect\citeauthoryear{Vanzella \etal}{2008}]{vanz08}
  Vanzella, E., \etal 2008, \aap, 478, 83
  
\bibitem[\protect\citeauthoryear{Vazdekis \etal}{1997}]{vaz97}
  Vazdekis, A., Peletier, R.~F., Beckman, J.~E. \& Casuso, E. 
  1997, \apjs, 111, 203

\bibitem[\protect\citeauthoryear{Vazdekis \etal}{2001}]{vaz01}
  Vazdekis, A., Kuntschner, H., Davies, R.~L., Arimoto, N., 
  Nakamura, O. \& Peletier, R. 2001, \apj, 551, 127
  
\bibitem[\protect\citeauthoryear{van~der~Marel \& van~Dokkum}{2007}]{vdm07}
  van~der~Marel, R.~P. \& van~Dokkum, P.~G. 2007, \apj, 668, 756

\bibitem[\protect\citeauthoryear{van~der~Wel \etal}{2004}]{vdwel04}
  van~der~Wel, A., Franx, M., van~Dokkum, P.~G. \& Rix, H.-W. 2004, \apj, 601, 5
  
\bibitem[\protect\citeauthoryear{van~der~Wel \etal}{2005}]{vdwel05}
  van der Wel, A., Franx, M., van~Dokkum, P.~G., Rix, H.-W., 
  Illingworth, G.~D. \& Rosati, P. 2005, \apj, 631, 145

\bibitem[\protect\citeauthoryear{van~der~Wel \etal}{2008}]{vdwel08}
  van der Wel, A., Holden, B.~P., Zirm, A.~W., Franx, M., Rettura, A., 
  Illingworth, G.~D. \& Ford, H.~C. 2008, \apj, 688, 48

\bibitem[\protect\citeauthoryear{van~Dokkum \etal}{1998}]{vdk98}
  van~Dokkum, P.~G., Franx, M., Kelson, D.~D. \& 
  Illingworth, G.~D. 1998, \apj, 504, 17
  
\bibitem[\protect\citeauthoryear{van~Dokkum \etal}{2000}]{vdk00}
  van~Dokkum, P.~G., Franx, M., Fabricant, D., Illingworth, G.~D. \& 
  Kelson, D.~D. 2000, \apj, 541, 951
  
\bibitem[\protect\citeauthoryear{van~Dokkum \& Ellis}{2003}]{vdk03}
  van~Dokkum, P.~G. \& Ellis, R.~S. 2003, \apj, 592, 53
  
\bibitem[\protect\citeauthoryear{van~Dokkum \etal}{2008}]{vdk08}
  van~Dokkum, P.~G., \etal 2008, \apj, 677, L5

\bibitem[\protect\citeauthoryear{van~Dokkum \etal}{2009}]{vdk09}
  van~Dokkum, P.~G., Kriek, M. \& Franx, M. 2009, \nat, in press, arXiv:0906.2778

\bibitem[\protect\citeauthoryear{Waddington \etal}{2002}]{wad02}  
  Waddington, I., \etal 2002, \mnras, 336, 1342

\bibitem[\protect\citeauthoryear{Whiley \etal}{2008}]{whil08}  
  Whiley, I~M., \etal 2008, \mnras, 387, 1253

\bibitem[\protect\citeauthoryear{Wiklind \etal}{2008}]{wik08}  
  Wiklind, T., Dickinson, M., Ferguson, H.~C., Giavalisco, M., Mobasher, B.,
  Grogin, N.~A. \& Panagia, N. 2008, \apj, 676, 781

\bibitem[\protect\citeauthoryear{Wirth \etal}{2004}]{tkrs04} 
  Wirth, G., et al., 2004, \aj, 127, 3121
  
\bibitem[\protect\citeauthoryear{Yamada \etal}{2006}]{yam06}
  Yamada, Y., Arimoto, N., Vazdekis, A. \& Peletier, R.~F.
  2006, \apj, 637, 200

\bibitem[\protect\citeauthoryear{Ziegler \& Bender}{1997}]{zieg97}
  Ziegler, B.~L. \& Bender, R. 1997, \mnras, 291, 527
  
\bibitem[\protect\citeauthoryear{Ziegler \etal}{1999}]{zieg99}
  Ziegler, B.~L., Saglia, R.~P., Bender, R. \& Belloni, P. 1999, A\&A, 346, 13

\end{thebibliography}

\clearpage

%\onecolumn

%%%%%%%%%%%%%%%%%%%%%%%%%%%%%%%%%%%%%%%%%%%%%%%%%%%%%%%%%
%%%%%%%%%%%%%%%%%%%%%%%%%%%%%%%%%%%%%%%%%%%%%%%%%%%%%%%%%
%%%%%%%%%%%%%%%%%    FIGURES      %%%%%%%%%%%%%%%%%%%%%%%
%%%%%%%%%%%%%%%%%%%%%%%%%%%%%%%%%%%%%%%%%%%%%%%%%%%%%%%%%
%%%%%%%%%%%%%%%%%%%%%%%%%%%%%%%%%%%%%%%%%%%%%%%%%%%%%%%%%

\end{document}